\documentclass[aps,preprint]{revtex4}%
\usepackage{amsfonts}
\usepackage{amsmath}
\usepackage{amssymb}
\usepackage{graphicx}%
\begin{document}

\title{Nonlinear electrodynamics at  cylindrical  ``cumulation''  fronts}

\author{F. Pegoraro $^1$\footnote{pegoraro@df.unipi.it},   S.V. Bulanov$^2$\footnote{Sergei.Bulanov@eli-beams.eu}}

\affiliation{$^1$  Physics Dept. 
Pisa University, largo Pontecorvo 3, 56127 Pisa, Italy
\\
National Research Council, National Institute of Optics,  via G. Moruzzi 1,  Pisa, Italy\\
 $^2$ Institute of Physics of the CAS, ELI-Beamlines Project, Na Slovance 2, 182 21 Prague, Czech Republic\\National Institutes for Quantum and Radiological Science and Technology (QST),
Kansai Photon Science Institute, 8--1--7 Umemidai, Kizugawa, Kyoto 619--0215, Japan}
\date{\today}


\begin{abstract}
Converging cylindrical electromagnetic  fields  in vacuum have been shown (E.I. Zababakhin, M.N. Nechaev, {\it Soviet Physics JETP},  {\bf 6},  345 (1958)) to exhibit  amplitude ``cumulation''. It was  found that the amplitude of self-similar waves increases without bounds  at finite distances from the
axis on the front of the fields reflected from the cylindrical axis. 
In the present paper we  propose to exploit  this cylindrical cumulation process as a possible new path towards  the generation  of ultra-strong electromagnetic  fields where nonlinear quantum electrodynamics (QED)  effects come into play.
We show that these effects, as described in the long wave-length limit within the framework of  the Euler 
Heisenberg Lagrangian,  induce a radius-dependent reduction  of the propagation speed of the cumulation front.   {Furthermore we compute  the $e^+$-$e^-$ pair production rate at the cumulation front  and show that the total number of pairs that are generated scales as the sixth power of the field amplitude.}

\end{abstract}
\maketitle

\section{Introduction}\label{intr}
As mentioned in Ref. \cite{peg} the ``recent developments in the generation of laser pulses with
ultra-high power (presently petawatt and progressing) have opened up  a new frontier  in plasma research  by making 
 it possible to obtain and to study ``mesoscopic''  amounts of
relativistic (ionised) matter  in compact-size experiments in the laboratory. 
This  will make it possible to investigate
in a  controlled environment    the nonlinear dynamics 
of collective relativistic systems, to enter the Quantum Electrodynamics  plasma regime,  and  to explore conditions that are of
interest for high energy astrophysics and beyond."\\ On the other hand, 
in the presence of  ultra-intense electromagnetic fields that approach the scale of the so called Schwinger field, i.e. of  the electric field  that corresponds to  an energy gain on a Compton length equal to the electron mass energy, vacuum itself behaves as a nonlinear medium where the electromagnetic waves induce  polarisation and magnetisation currents. 
 As recalled e.g. in Ref. \cite{hod}  in  ``classical electrodynamics electromagnetic waves do not interact in  vacuum. On the contrary, in QED photon-photon scattering can take
place in vacuum via the generation of virtual electron-positron pairs. This interaction gives rise to vacuum polarisation and birefringence,
to the Lamb shift, to a modification of the Coulomb field, and to many other phenomena \cite{BLP}'' {(see Refs.\cite{piazza,king,rend})}.

The electromagnetic fields in present laser pulses  do not  approach the magnitude of the Schwinger field
($E_s = m_e^2 c^3/{e\hbar} = 1.32 \times 10^{18} V/m$ where the symbols have their standard meaning),
but the nonlinear
 properties  of relativistic plasmas, and in particular the so called relativistic  flying mirrors { \cite{bulan,MTB,BEK,BulR} can be used  to significantly enhance  and focus the electromagnetic energy  of  presently available laser pulses. 
 As noted in \cite{atto}, ``the measurement and control of sub-cycle field evolution of few-cycle
light have opened the door to a radically new approach to exploring and controlling processes of the microcosm.''

 An alternative approach is to search for geometrical configurations and pulse shapes that can lead to a local enhancement  of the pulse field intensity through the process of amplitude cumulation, i.e. to a    local  formal divergence of the field intensity through a mechanism of constructive interference. In fact, converging cylindrical electromagnetic  fields  in vacuum have been shown \cite{cumul} to exhibit  amplitude cumulation. It was  found that the amplitude of self-similar waves increases without bounds  at finite distances from the
axis on the front of the fields reflected from the cylindrical axis.  It was  remarked  that this cumulation  process is a different phenomenon with respect to the $r^{-1/2}$  dependence close to the  cylinder axis,  $r$ being  the distance from the axis,  that follows simply from the Poynting  flux conservation in a cylindrical configuration.   The amplitude  cumulation  is produced  by the constructive interference  that is induced because the  self-similar electromagnetic pulse  has a fully coherent spectrum which  extends over the whole frequency range and  decays for large frequencies as the inverse of the frequency square-root.
The effects of the pulse  geometry and coherence  combine in such a way that after the  reflection from the cylindrical  axis a wave front is formed that propagates outwards at the speed of light. At this front the field amplitude  formally diverges.  In Ref. \cite{cumul}   the important  fact that this singularity is not limited to the axis but sweeps a wide area was  emphasised.

An obvious  problem when  applying the amplitude cumulation  mechanism to the creation of ultra intense electromagnetic fields arises as  to  how such a self-similar  pulse can be prepared and how stable it is to small deviations in its spectral composition and in its phase coherence. In Ref.\cite{cumul}  the point was made that   a self-similar  solution of the Maxwell's equations ``describes the limiting behaviour of a field close to the axis and close to the time of focusing''.  On the other hand, in the  context  of the implosion of cylindrical liners,
this   idealised solution has been explicitly criticised in Ref.\cite{critic} on the point that it does not satisfy the 
``the boundary condition imposed by magnetic flux conservation  for field compression by an ideally conducting cylinder'', i.e. for the configuration  for which it was initially constructed.

In the present paper we do not directly address these points but exploit the fact that in a cylindrical electromagnetic wave the Lorentz invariant ${\bf E}^2 - {\bf B}^2$ {(also known as the Poincar\'e invariant of the electromagnetic field)} does not vanish,  as would be  the case for plane electromagnetic  waves.  This fact, together with the large fields produced at  the cumulation front, can make this configuration interesting for the study of QED effects in vacuum within the framework described by the Euler-Heisenberg \cite{EH} Lagrangian. In fact a converging cylindrical configuration can be seen as a   limiting case combining   the  converging multi-light-beam approach  adopted e.g. in Ref.\cite{bulanino} and, in view of  the reflection at  the magnetic axis, the counter propagating  laser  pulse approach (for a recent investigation of this latter approach see Ref. \cite{hod}).
{Similarly,  these combined effects can  be exploited in order to enhance   the production of $e^+$-$e^-$ pairs at the cumulation front.

This article is organised as follows. In Sec.\ref{EHaction}   the reduced Lagrangian densities for $s$- and for $p$-polarised fields  in cylindrical geometry are derived  from the  general electromagnetic  vacuum  action functional  that includes   the long wave-length QED corrections to classical  electrodynamics.  In Sec.\ref{classical}
the classical limit is considered and the self-similar solutions are explicitly constructed and found  to  involve Elliptic integrals of the self-similar variable $\tau = ct/r$. It is shown that at the cumulation front both the electromagnetic fields and the Lorentz invariants diverge logarithmically.  The  spectrum of the self-similar solutions is computed and shown to decay as the inverse square root of the frequency for large frequencies. In Sec.\ref{qedcorr} the  QED corrections are computed explicitly for the $s$-polarised electromagnetic fields and interpreted in terms of  a reduction of the wave propagation speed.  
{ In Sec.\ref{pair} the production rate of electron-positron pairs  in the region near the cumulation front and the total number of pairs  created by a self-similar pulse are computed.}
Finally, in Sec.\ref{concl} we formulate our conclusions.}

\section{Euler Heisenberg Lagrangian density}\label{EHaction}

In the long wave-length limit the electromagnetic  action functional  ${\cal S} $ in vacuum  that includes   the QED corrections \cite{EH,BLP} to classical  electrodynamics  can be expressed  as 
 \begin{equation}
 {\cal S} = \int d^3x\,  d t  \,  \mathcal{L}, \quad {\rm with} \quad \mathcal{L}=\mathcal{L}_{0}+\mathcal{L}', \label{action}\end{equation}
 where
\begin{equation}
\mathcal{L}_{0}=-\frac{1}{16\pi}F_{\mu \nu}F^{\mu \nu}
\end{equation}
 is the  Lagrangian density of  classical electrodynamics in vacuum 
 and  $\mathcal{L}'$ is the Heisenberg--Euler Lagrangian  density.  
 Here   $F_{\mu \nu}$ is the electromagnetic field tensor 
\begin{equation}
F_{\mu \nu}=\partial_{\mu} A_{\nu}-\partial_{\nu} A_{\mu},
\end{equation}
with $A_{\mu}$  the 4-vector of the electromagnetic field
and $\mu=0,1,2,3$. Here and below we assume  summation over repeated indices  and adopt natural units setting $c = \hbar = 1$.

In the weak field approximation (see e.g.  \cite{HeHe}) the Heisenberg--Euler Lagrangian  density  $\mathcal{L}'$   can be written as 
\begin{equation}
\mathcal{L}'=\frac{\kappa}{4}\left[{\mathfrak F}^2 
+ \frac{7}{4} {\mathfrak G}^2 +{\frac{2}{7} } {\mathfrak F} \left(  {\mathfrak F}^2 +\frac{13}{16}{\mathfrak G}^2  \right) \right]+...
\label{eq:mathcalL}
\end{equation}
 with $\kappa=e^4/360 \pi^2 {m_e}^4$ and $  {\mathfrak F}$ and{ ${\mathfrak G}$}
 the Poincar\'e invariants 
\begin{equation} {\mathfrak F}=F_{\mu \nu}F^{\mu \nu} \quad {\rm and} \quad  {\mathfrak G}=F_{\mu \nu}\tilde F^{\mu \nu},
\end{equation} and  $\tilde F^{\mu \nu}=\varepsilon^{\mu \nu \rho \sigma}F_{\rho \sigma}$ is the dual electromagnetic tensor 
with $\varepsilon^{\mu \nu \rho \sigma}$  the Levi-Civita symbol in four dimensions. In the above equations, and in the following sections  unless explicitly stated, the electromagnetic fields  are normalised on the Schwinger field.
In the Lagrangian density  (\ref{eq:mathcalL}) the first two terms on the right hand side   and the last two correspond respectively to  four and to six photon interaction, respectively.
In the following, for the sake of simplicity, {we will  retain only the four photon interaction term}.

\subsection{Cylindrical waves}

Here we address  the  propagation of   converging cylindrical electromagnetic waves with either $s$-type or $p$-type polarisation.  We will consider the two polarisations separately, aside for Sec.(\ref{classical}) where  the  cylindrical wave   equations  are derived by neglecting the Euler-Heisenberg correction to the classical  Lagrangian density ${\cal L}_o$  and thus obey the  superposition principle. This simply amounts to a simplification of the analysis as in this 
 case the invariant ${\mathfrak G}$ that would couple the two polarisations in the Euler-Heisenberg Lagrangian density  vanishes identically. 
 In physical terms  this amounts to including the effect of photon-photon scattering  while disregarding the effect of vacuum induced birefringence. Furthermore we assume translational invariance along  $z$, i.e. along the axis direction, and  azimuthal  invariance along the angle  $\varphi$. 

The $s$-type   waves can be described  in a transverse gauge by a  vector potential with a single  component, ${\bf A}=A_s (r, t) \, {\bf e}_z$, where  ${\bf e}_z$ is 
the unit vector along the $z$ axis.  Similarly, the $p$-type   waves can be described   by a  vector potential with  a single  component, ${\bf A}=A_p (r,t)\,  {\bf e}_\varphi$, with ${\bf e}_\varphi$
the unit vector along the azimuthal direction.
Factoring  the two invariance directions out of the action functional, we obtain with obvious notation for the two polarisations separately
 \begin{equation}
 {\cal S}_s \propto \int r dr \,  d t  \,  \mathcal{L}_s \left( A_s (r, t) \right), \quad   {\cal S}_p \propto \int r dr \,  d t  \,\mathcal{L}_p \left( A_p (r, t) \right)     \label{actioncyl}
 \end{equation}
where 
\begin{align}  \label{Ls}
& 
\mathcal{L}_s \left( A_s (r, t) \right) = \frac{1}{8\pi} \left[E_z^2 - B_\varphi^2+ \epsilon (E_z^2 - B_\varphi^2)^2 \right] = 
\\  
& \frac{1}{8\pi} \left[ \left(\frac{\partial A_z}{\partial t}\right)^2 - \left( \frac{\partial A_z}{\partial r}\right)^2  
+ \epsilon \left[   \left(\frac{\partial A_z}{\partial t}\right)^2 - \left( \frac{\partial A_z}{\partial r}\right)^2 \right]^2 \right]
\nonumber 
\end{align} 
\begin{align}  \label{Lp}
& 
\mathcal{L}_p \left( A_s (r, t) \right) = \frac{1}{8\pi} \left[E_\varphi^2 - B_z^2+ \epsilon (E_\varphi^2 - B_z^2)^2 \right] = 
\\ 
& \frac{1}{8\pi} \left[ \left(\frac{\partial A_\varphi}{\partial t}\right)^2 - \left( \frac{1}{r} \frac{\partial \,( r A_\varphi)}{\partial r}\right)^2  + \epsilon \left[   \left(\frac{\partial A_\varphi}{\partial t}\right)^2 - \left(  \frac{1}{r} \frac{\partial \,  (r A_\varphi)}{\partial r}\right)^2 \right]^2 \right]
\nonumber \end{align} 
with $\epsilon = e^2/(4 \pi) = \alpha /(4 \pi)$ where 
{$\alpha=e^2/\hbar c \approx 1/137$} is the fine structure constant. 

\section{Classical electrodynamics limit, $\epsilon \to 0$}\label{classical}

Varying the Action functional (\ref{actioncyl}) with respect to the two components of the vector potential and expressing the resulting equations in the limit $\epsilon \to 0$  in terms of the electromagnetic fields, we obtain the field equations
\begin{equation}\label{01=}
\frac{\partial{E}_z}{\partial t} =   \frac{1}{ r} \frac{\partial (rB_\varphi)}{\partial r}, 
\quad {\rm with} \quad \frac{\partial{B}_\varphi}{\partial t} =   \frac{\partial E_z}{\partial r} , \end{equation}
 for the $s$-polarisation and 
\begin{equation}\label{02=} 
\frac{\partial{E}_\varphi}{\partial t} =  - \frac{\partial B_z}{\partial r},
\quad {\rm with } \quad   \frac{\partial{B}_z}{\partial t} =  -  \frac{1}{ r} \frac{\partial (rE_\varphi)}{\partial r},
\end{equation}
 for the $p$-polarisation. {The two polarisations are related by the symmetry transformation 
${\bf B} \to {\bf E}$ and ${\bf E} \to -{\bf B}$ which is characteristic of classical electrodynamics in vacuum.}
\, Eqs.(\ref{01=}, \ref{02=}) lead to  the cylindrical wave equations 
\begin{equation}\label{2=}
\frac{\partial^2{E}_z}{\partial t^2} -  \frac{1 }{r}\ \frac{\partial }{\partial r} \left(  r \frac{\partial {E}_z}{\partial r} \right)=0,
\qquad \frac{\partial^2{B}_\varphi}{\partial t^2} - \frac{\partial }{\partial r} \left( \frac{1 }{r}\frac{\partial (r {B}_\varphi)}{\partial r} \right)=0,\end{equation}
for the $s$-polarisation and to the corresponding one with ${\bf E}$ and ${\bf B}$ interchanged for the $p$-polarisation.

\subsection{Reduced fields} \label{reduced}

In  cylindrical geometry it is    convenient   to introduce the reduced electromagnetic  fields 
\begin{equation}\label{3=}
 {E}_{z ,\varphi}(r,t) = \frac{e_{s,p}(r,t) }{r^{1/2}},  \qquad  {B}_{z,\varphi}  (r,t)= \frac{ b_{s,p}(r,t) }{r^{1/2}} ,
\end{equation}
where $r$  and $t$ are now dimensionless space-time coordinates normalised on a  spatial reference scale $r_o$.\,
Then for the $s$-polarisation   we obtain 
\begin{equation}\label{1=}
 \frac{\partial{e_s}}{\partial t} =  \frac{\partial b_s}{\partial r}  + \frac{b_s}{2r}, \qquad  \frac{\partial{b_s}}{\partial t} =   \frac{\partial e_s}{\partial r} - \frac{e_s}{2r},
\end{equation}
and for the $p$-polarisation  
\begin{equation}\label{4=}
 \frac{\partial{e_p}}{\partial t} =  - \frac{\partial b_p}{\partial r} + \frac{b_p}{2r}, \qquad  \frac{\partial{b_p}}{\partial t} =  - \frac{\partial e_p}{\partial r} - \frac{e_p}{2r}.
\end{equation}
From Eqs.(\ref{1=},\ref{4=})  we obtain the  reduced form of the Poynting flux  and the radial  electromagnetic momentum density  equations    for each polarisation separately 
\begin{equation}\label{5=}
\frac{\partial  (e_s^2 + b_s^2)/2 } {\partial t}  -  \frac{\partial  (e_s b_s) }{\partial r} = 0, \qquad
 \frac{\partial  (e_p^2 + b_p^2 )/2 } {\partial t}  +  \frac{\partial  (e_p b_p) }{\partial r} = 0,
\end{equation}
\begin{align}\label{6=} &\frac{\partial  (e_s b_s)} {\partial t}  -  \frac{\partial  (e_s^2 + b_s^2)/2  }{\partial r} =
\frac{b_s^2 - e_s^2}{2 r} , \\ &\frac{\partial  (e_p b_p)} {\partial t}  +  \frac{\partial  (e_p^2 + b_p^2)/2  }{\partial r} =
 \frac{b_p^2 - e_p^2}{2 r} . \nonumber \end{align}
In addition the mixed polarisation equations hold  
 \begin{align}\label{6+=}  & \frac{\partial  (e_p b_s  -  e_s b_p)  }{\partial t} -  \frac{\partial  (e_s e_p - b_s b_p)} {\partial r}  = 0,
\\ & \frac{\partial  (e_s e_p - b_s b_p)} {\partial t}  -  \frac{\partial  (e_p b_s  -  e_s b_p)  }{\partial r} =
2 \, \frac{e_p b_s + e_s b_p}{ 2r} . \nonumber \end{align}
where $ e_s b_p  + e_p b_s  = r \, {\bf E}\cdot {\bf B}=r\, { {\mathfrak G}}$. 
Note that the ``source terms'' in the r.h.s. of Eqs. (\ref{6=}) and on the second of  Eqs. (\ref{6+=})  are the terms that determine the magnitude of the Euler-Heisenberg contribution  relative to the classical part  in the electromagnetic Lagrangian density.

\subsection{Self-similar fields} \label{s-sim}

Following Ref.\cite{cumul}, we look for self-similar solutions of Eqs.(\ref{1=},\ref{4=}) by assuming that the fields $e_{s,p}(r,t) $ and $b_{s,p}(r,t) $ 
depend on the single variable   $\tau = t/r$. We obtain
\begin{align}\label{7=} 
&\frac{ d e_s}{d \tau}= -\tau \frac{ d b_s }{d\tau}  + \frac{b_s}{2}, \quad  \frac{d b_s}{d\tau} = -\tau {\frac{d e_s}{d \tau}}   -\frac{ e_s}{2}  \nonumber  \\
&\frac{ d e_p}{d \tau}= \tau \frac{ d b_p }{d\tau}  + \frac{b_p}{2}, \quad  \frac{d b_p}{d\tau} ={ \tau \frac{d e_p}{d \tau}}   -\frac{ e_p}{2} . \end{align}
{We set the time origin such that  $\tau <0$ corresponds to the converging part of the solution and $\tau >0$  to the diverging one.  Consistently,  we impose that on the converging part   $e_s$ and $b_s$ have the same sign while $e_p$ and $b_p$ have opposite signs, as implied by propagation towards the cylinder axis.\\
In particular at $\tau = -1$  (i.e. at $r = |t|$, with $t<0$)  from Eqs.(\ref{7=}) we obtain 
\begin{equation}\label{7==} 
e_s(\tau = -1) =  b_s(\tau = -1), \quad e_p(\tau = -1) =  - b_p(\tau = -1). \end{equation}}
Differentiating Eqs.(\ref{7=})  with respect to $\tau$ we obtain for the pairs $(e_s,\, b_p)$ and $(e_p,\, b_s)$ the second order ordinary differential equations 
\begin{equation}\label{8=} 
{
\frac{d } {d \tau} \left[(1-\tau^2)  \frac{d } {\partial \tau}\right]  (e_s,\, b_p)=\frac{1}{4} \, (e_s,\, b_p) 
}, 
\end{equation}
\begin{equation}\label{8+} 
{ \frac{d } {d \tau} \left[(1-\tau^2)  \frac{d } {\partial \tau}\right]  (e_p,\, b_s) = - \frac{3}{4}\, (e_p,\, b_s ).}
 \end{equation}
Equations (\ref{8=},\ref{8+})  are singular at $\tau = \pm 1$ where  a local analysis gives  the two independent solutions in the form
\begin{equation}\label{8+-} 
C_\pm \ln{|(\tau \mp 1) |}+.... \quad {\rm and} \quad  D_\pm +....   \,  ,
 \end{equation}
 where $C_\pm$ and $D_\pm $ are constants with the index $\pm$ referring to $\tau = \pm 1$ respectively and only the leading term  of the local expansions is shown. \\ In the following we require  that the converging part of the solution for the electromagnetic fields  be regular and  thus set  $C_- = 0$,  i.e. we impose that  the solution have no logarithmic singularity at $\tau = -1$.
\\
Equations (\ref{8=},\ref{8+})  are Legendre equations of index $\nu = -1/2$ and $\nu = 1/2$ respectively. Following Ref.\cite{NIST} we write their solutions in the interval $-1 \le \tau\le1$ in terms of elliptic integrals  as
\begin{align}\label{KQn1} 
& ~ e_s (\tau) =\frac{2}{\pi}\, {\rm Q}_{-1/2}(\tau) = \frac{2}{\pi}\, K(\sqrt{(1 + \tau)/2}),  \\
& e_p (\tau) =  -\frac{2}{\pi}\, {\rm Q}_{1/2}(\tau) = -\frac{2}{\pi}\,  K(\sqrt{(1 + \tau)/2})  
 + \frac{4}{\pi}\, E(\sqrt{(1 + \tau)/2})  ,\nonumber  \end{align}
where ${\rm Q}_{\pm 1/2}(\tau)$  are Legendre functions (see Ref.\cite{NIST}) and   $K(\sqrt{(1 + \tau)/2})$ and $E(\sqrt{(1 + \tau)/2})$  are complete elliptic integrals {of the first and second kind, respectively}.  In accordance with the  regularity condition at $\tau =-1$, in Eqs.(\ref{KQn1}) 
 we have imposed $e_s(\tau =-1) =e_p(\tau =-1) = 1$.
 For $\tau \to 1$  the solutions (\ref{KQn1}) display a  logarithmic singularity as $E \to 1 $ while  $K \sim  - (1/2) \ln{(1-\tau)} \to  +\infty$.
 \\
The solutions of Eqs.(\ref{8=},\ref{8+})  in the interval $1 \le \tau$  that vanish for $\tau \to \infty$, i.e. that vanish on the  cylinder axis, can be written as 
\begin{align}\label{KQninfty} 
& ~ e_s (\tau) = \frac{1}{\pi^{1/2}} \, {\mathbf {\cal Q}}_{-1/2}(\tau) = \frac{2}{\pi} \,  
\frac{ K( 1/(\tau + \sqrt{\tau^2 -1}))}{\sqrt{ \tau + (\tau^2 -1)}}, \nonumber  \\
& e_p (\tau) =  - \frac{1}{\pi^{1/2}}\, {\mathbf {\cal Q}}_{1/2}(\tau) = - \frac{2}{\pi} \, \frac{ K( 1/(\tau + \sqrt{\tau^2 -1}))}{\sqrt{ \tau + (\tau^2 -1)}} \nonumber \\
& \qquad \qquad  \qquad \qquad \quad \qquad  + \frac{4}{\pi}  \frac{ E( 1/(\tau + \sqrt{\tau^2 -1}))}{\sqrt{ \tau + (\tau^2 -1)}} ,\end{align}
where ${\mathbf {\cal Q}}_{\pm 1/2}(\tau)$  are Legendre functions \cite{NIST}  and the coefficient has been fixed  by requiring that the electromagnetic fields be continuous at $\tau = 1$.
\\
Note in passing that, when inserted into Eqs.(\ref{3=}) these solutions lead to electric and magnetic fields that are regular at $r=0$ for $t>0$ and that depend on time as $t ^{-1/2}$.
\\The expressions for the magnetic field components $b_s$ and $b_p$ are obtained by inserting $b_s$ for $e_p$ and $- b_p$  for $e_s$ in Eqs.(\ref{KQn1},\ref{KQninfty}).

The logarithmic   singularity at $\tau = 1$  corresponds  to the  cumulation  process  first identified in   Ref.\cite{cumul}. It  occurs at a  finite distance  from the axis, at the front of the reflected  fields, and  propagates outwards at the speed of light.
\\
The coefficients of the logarithmic singularity at $\tau = 1$ in the electric and magnetic fields   are related to each other  by  the leading order terms in Eqs.(\ref{7=}) and are thus either equal or  equal and opposite. However, for each of the polarisation,  the next order terms for  the electric and magnetic field expansion  differ:  thus in the terms   $b_{s,p}^2 - e_{s,p}^2$  and $e_{s}\,b_{p} + e_{p}\,b_{s}$  the contributions
proportional to the  logarithm squared  cancel  out while the linear ones in the logarithm do not.\,
More explicitly for $-1 < \tau <1$  from Eqs.(\ref{KQn1}), using the  relationships between $b_s$ and $e_p$ and between  $b_p$ and $e_s$ mentioned above,   and setting  $e_s(\tau =1) = A_s,$ and  $e_p(\tau = 1) =A_p$, we find 
\begin{align}\label{Inv1} 
&e_s^2 - b_s^2 =  
\left[\frac{16 A^2_s}{\pi^2}\right] {\cal E}, \quad  e_p^2 - b_p^2 =   -\left[\frac{16 A^2_p}{\pi^2}\right]  {\cal E}, \quad  e_p b_s + e_s b_p = -\left[\frac{16 A_p B_p}{\pi^2}\right] {\cal E}, \nonumber\\
& {\rm with} \quad {\cal E}  =  K(\sqrt{(1 + \tau)/2}) E(\sqrt{(1 + \tau)/2}) -  E^2(\sqrt{(1 + \tau)/2}) \ge  0 ,   \end{align}
which diverges as $ - (1/2) \ln{(1-\tau)}$ for $\tau \to 1,$ see Fig.(\ref{plot1}) \,
Note the opposite sign  of the invariants ${\bf E}^2 - {\bf B}^2$ in  Eq.(\ref{Inv1}) between the $s$ and the $p$ polarisations.
\begin{figure}[h!]
\centering
\includegraphics[height=6cm]{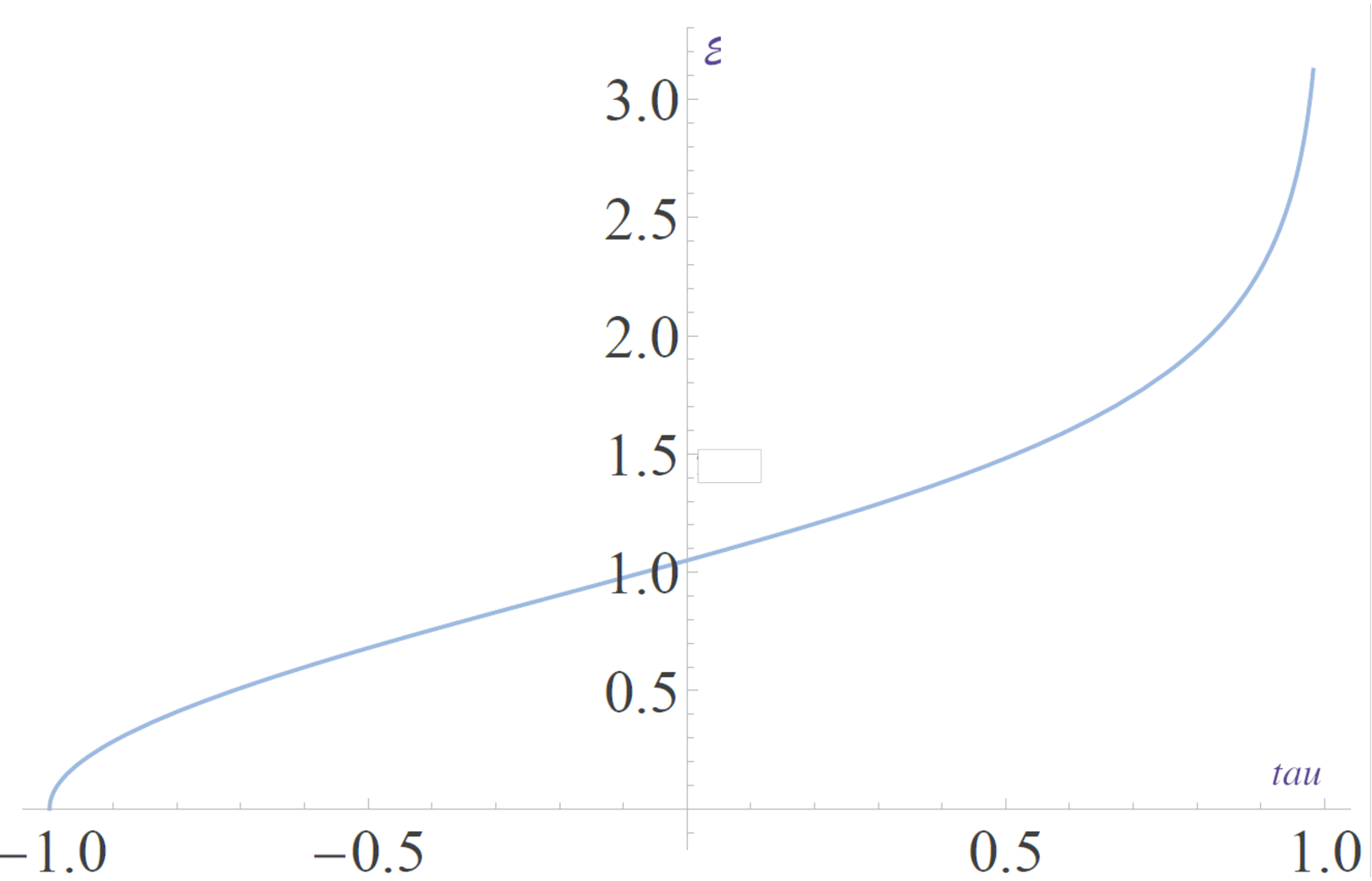}
\caption{Plot of ${\cal E} =  K(\sqrt{(1 + \tau)/2}) E(\sqrt{(1 + \tau)/2}) -  E^2(\sqrt{(1 + \tau)/2}) $.   }
\label{plot1}
\end{figure}
Corresponding formulae can be obtained from Eqs.(\ref{KQninfty}) for $\tau>1$.\\
The cancellation of the leading order terms in the Lorentz invariants $e_s^2 - b_s^2 $,  
$e_p^2 - b_p^2$ and $ e_p b_s + e_s b_p $ can be also seen without resorting to the explicit solutions (\ref{KQn1},\ref{KQninfty})  by rewriting Eqs.(\ref{5=}, \ref{6=}) in terms of the self-similar fields.  We obtain
\begin{align}\label{9=}
& \frac{d  (e_{s,p}^2 + b_{s,p}^2)/2 } {d \tau }  = (\mp) \, \tau   \frac{d  (e_{s,p}\,b_{s,p}) }{d \tau}, \\ &
\frac{d  (e_{s,p}\, b_{s,p})} {d \tau }  = (\mp)\, \tau   \frac{d  (e_{s,p}^2 + b_{s,p}^2)/2  }{d \tau} +\frac{b_{s,p}^2 - e_{s,p}^2}{2},\nonumber 
\end{align}
with $(\mp) = -$ for the $s$ polarisation and $+$ for the $p$-polarisation,
which give  for both polarisations the following equation for the reduced Poynting flux
\begin{equation}\label{10=}
(1 -\tau^2) \frac{d (e_{s,p}\,b_{s,p})}{d \tau }  = \frac{b_{s,p}^2 - e_{s,p}^2}{2}.
\end{equation}
Similarly from Eqs.(\ref{6+=}) we obtain for  the mixed polarisation case 
 \begin{align}\label{9+=} 
  & { \frac{d  (e_p b_s  -  e_s b_p)  }{d \tau } + \tau  \frac{d  (e_s e_p - b_s b_p)} {d \tau}  = 0, }
 \\&
{ \frac{d  (e_s e_p - b_s b_p)} {d \tau}  + \tau  \frac{d  (e_p b_s  -  e_s b_p)  }{d \tau } =
e_p b_s + e_s b_p },\nonumber 
\end{align} 
which give
\begin{equation}\label{10+=}
(1 -\tau^2) \frac{d (e_s e_p - b_s b_p)}{d \tau }  = e_p b_s + e_s b_p.
\end{equation}
Inserting the local  expansion (\ref{8+-})   into Eqs.(\ref{10=},\ref{10+=}), in the  neighbourhood of $\tau = 1$  we recover 
$b_s^2 - e_s^2  \propto  \ln{|1 -\tau|} , \,  b_p^2 - e_p^2  \propto -  \ln{|1 -\tau|},\,  e_p b_s + e_s b_p  \propto -  \ln{|1 -\tau|}$. 

\subsection{Frequency spectrum of the  self-similar fields} \label{spectrum}

Let us consider the cylindrical wave equations (\ref{2=}) with 
\begin{equation}\label{s2}
(E_z(r,t), \, B_z(r,t))= \frac{1}{(2\pi)^{1/2}}\int_{0}^{+\infty} d \omega \, \, ({\hat E}(\omega,r), {\hat B}(\omega,r)) \, \,  \exp{(-i \omega t)}\,  +c.c.
\end{equation}
Then 
\begin{equation}\label{s3}
 \frac{1 }{r}\ \frac{\partial }{\partial r} \left(  r \frac{\partial ({\hat E}, {\hat B})}{\partial r} \right)=- \frac{\omega^2 ({\hat E}, {\hat B}) }{c^2} .\end{equation}
We write ${\hat E}(\omega,r) $ (and  analogously for ${\hat B}(\omega,r)$) in the form
\begin{equation}\label{s4}
{\hat E}(\omega,r) = {\cal U}(\omega)\,  { {\bar E}}( \omega r/c) ,
\end{equation}
where 
${ {\bar E}}( \omega r/c)$ denotes  a combination, depending on the initial conditions,  of Bessel functions of index $o$.  
Then Eqs.(\ref{s2},\ref{s4}) together with the self-similarity condition can be written (in dimensional units) as 
\begin{equation}\label{s5}
e_s(ct/r) = \frac{r^{1/2}}{(2\pi)^{1/2}}\int_{0}^{+\infty} d \omega \, \,  {\cal U}(\omega)\,  { {\bar E}}  ( \omega r/c)\,  \exp{(-i \omega t)} + c.c,
\end{equation}
i.e., 
\begin{equation}\label{s7}
e_s(ct/r) = \frac{c}{(2\pi)^{1/2}}\int_{0}^{+\infty} d (\omega r/c) \, \, r^{-1/2} {\cal U}(\omega)\,  { {\bar E}}  ( \omega r/c)\,  \exp{[(-i \omega r/c) (ct)/r]} +c.c. .
\end{equation}
Consistency between the r.h.s. and the l.h.s. of Eq.(\ref{s7}) requires  that 
\begin{equation}\label{s8}
 r^{-1/2} {\cal U}(\omega)\, =  Const \,~\,(\omega r/c)^{-1/2},\qquad 
 {\rm which\,\, implies} 
  \quad  {\cal U}(\omega)\ \propto \omega^{-1/2},
\end{equation}
which  corresponds to a slow decay of the frequency spectrum for $\omega \to \infty$ with all the frequency components being in phase.

{We  observe that the singularity at $ \tau = 1$ can be re-derived by inserting the large $\omega r/c$ behaviour  of the Hankel function 
of the first kind for  ${ {\bar E}} $ into the integrand in Eq.(\ref{s7}) and by evaluating the integral explicitly. 
\\
This procedure  allows  us  to compute the modification of the logarithmic singularity if we impose a cut-off in the frequency spectrum while preserving the spectrum coherence. If, for example, we impose a Gaussian cut-off  by modifying Eq.(\ref{s8}) and setting  
\begin{equation}\label{s9} r^{-1/2} {\cal U}(\omega)\, =  Const \, \exp{[-(\omega/\omega_{max})^2]}~\,(\omega r/c)^{-1/2}.\end{equation}
we find that  $e_s(\tau = 1) \propto \ln{(\omega_{max} r/c)^2}$. }

\section{QED corrections to the propagation of the cumulation front}\label{qedcorr}

Varying the Action functional (\ref{actioncyl}) with respect to the two components of the vector potential  separately and expressing the resulting equations in terms of the electromagnetic fields, we obtain  for $\epsilon \not= 0$ the field equations

\begin{align}\label{EH1=}
&\frac{\partial{E}_z}{\partial t}  + 2\epsilon \frac{\partial }{\partial t} \,  \left [E_z^3 -  E_z B_\varphi^2\right] =   \frac{1}{ r} \frac{\partial (rB_\varphi)}{\partial r} +
 2 \epsilon \frac{1}{ r} \frac{\partial }{\partial r}\left[ r(B_\varphi E_z^2 - B_\varphi^3)\right], 
\nonumber \\&{\rm with} \quad \frac{\partial{B}_\varphi}{\partial t} =   \frac{\partial E_z}{\partial r} \end{align}
 for the $s$-polarisation  and  for the $p$-polarisation
\begin{align}\label{EH2=} &
\frac{\partial{E}_\varphi}{\partial t}   + 2\epsilon \frac{\partial }{\partial t} \,  \left [E_\varphi^3 -  E_\varphi B_z^2\right] =  - \frac{\partial B_z}{\partial r} - 2 \epsilon 
\frac{\partial }{\partial r} \,  \left [B_z E_\varphi^2 -  B_z^3\right]  , \nonumber 
\\ & {\rm with } \quad   \frac{\partial{B}_z}{\partial t} =  -  \frac{1}{ r} \frac{\partial (rE_\varphi)}{\partial r}.
\end{align}
In the following we will 
  examine the effect of the Euler-Heisenberg corrections  on the self-similar solutions described in Sec.\ref{s-sim}  
  by adopting a perturbative procedure in $\epsilon/r$, where the additional geometrical factor $1/r$ accounts for the fact that the nonlinear terms grow as $1/r$  in comparison with the linear terms as the cylinder axis is approached.  Furthermore, 
  in order not to duplicate the derivation,  we will only refer explicitly to the $s$ polarised case. \, Following  Eq.(\ref{3=}), 
we  use the reduced fields $e_s(r,t)$ and $b_s(r,t)$ which we expand as
  \begin{equation}  \label{pert00}
e_s (r,t)  = e_{s0}(\tau) + \frac{\epsilon}{r}  \, e_{s1}(\tau)  +.. ,\quad  b_s = b_{s0}(\tau) + \frac{\epsilon}{r} \, b_{s1}(\tau) + ...., 
\end{equation}
where $ e_{s0}(\tau)$  and $ b_{s0}(\tau)$   are given by Eqs.(\ref{KQn1}, \ref{KQninfty}) with $b_{s0} (\tau)= e_{p0}(\tau)$.  Then to first order in $\epsilon/r$ we obtain 
  \begin{align}\label{EH3=}
& \frac{d {e_{s1}}}{d\tau} + \tau \frac{d b_{s1}}{d \tau} 
+\frac{b_{s1}}{2}  = 
\\ & -2   \frac{d{(e_{s0}^3 - e_{s0} b_{s0}^2})}{d\tau}  
- 2\tau  \frac{d{(e_{s0}^2 b_s -   b_{s0}^3})}{d \tau} 
 - (e_{s0}^2 b_{s0} -   b_{s0}^3)  , 
\nonumber \\&
\frac{d b_{s1}}{d \tau} + \tau  \frac{d e_{s1}}{d \tau}  +  \frac{ 3\, e_{s1}}{2} = 0, \nonumber
\end{align}
where a common  factor $1/r^2$ as been  removed.
 Equations (\ref{EH3=}) can be recast in the form
\begin{align}&  \label{pert2} 
\frac{d }{d \tau} (1-\tau^2)  \frac{d e_{s1}}{d \tau}  - 2 \tau  \frac{d e_{s1}}{d \tau} - \frac{9}{4} e_{s1} = - \frac {d{\cal  Q}_0} {d \tau},
 \nonumber \\
& 
\frac{d }{d \tau} (1-\tau^2)  \frac{d b_{s1}}{d \tau}  - 2 \tau  \frac{d b_{s1}}{d \tau}  -\frac{5\, b_{s1}}{4} =  \tau \frac{d {{\cal  Q}_0}}{d\tau}  + \frac{5}{2} {\cal Q}_0%
  \end{align}
   to be solved in the domain $-1 < \tau < +\infty$.
Here 
\begin{equation}\label{pert1b}
{{\cal  Q}_0}(e_{s0},b_{s0}) = 
2  \frac{d (e_{s0}^3- e_{s0} b_{s0}^2)}{d \tau} + 2 \tau  \frac{d ( e_{s0}^2 b_{s0} -   b_{s0}^3)}{d \tau} + (e_{s0}^2 b_{s0} -   b_{s0}^3). \end{equation}
In order to limit  the number of required algebraic manipulations, in the following  it will suffice to present the solutions of Eq.(\ref{pert2}) explicitly only in the interval $-1 < \tau < 1$, i.e. in front of the cumulation singularity.   \\
Using the expressions in  in Eqs.(\ref{KQn1})  we find, see Fig.(\ref{plotS}),
\begin{align}\label{explic}  &  ~\qquad \qquad  {{\cal  Q}_0} (\tau)  = 
- \left[(16/\pi^3)/\left[  (1+\tau) (\sqrt{(1 + \tau)/2}-1)\right] \right]\\
&  \left[\,  (\, \tau\, (10 \sqrt{(1 + \tau)/2} -7)+ 4 \sqrt{(1 + \tau)/2} -5 ) \, E(\sqrt{(1 + \tau)/2})^3    \right.  \nonumber  \\   
&  \left.  +(\tau((13 -15 \sqrt{(1 + \tau)/2})- 5 \sqrt{(1 + \tau)/2}+7 ) \,  \right.    \nonumber  \\   
&  \left. ~\qquad \qquad  ~\qquad \qquad  ~\qquad \quad K(\sqrt{(1 + \tau)/2}) \, E(\sqrt{(1 + \tau)/2})^2     \right.   \nonumber   \\ 
&  \left.   + (1+7 \tau) ) (\sqrt{(1 + \tau)/2} -1)  \, K(\sqrt{(1 + \tau)/2})^2  \, E(\sqrt{(1 + \tau)/2})   \right.   \nonumber   \\
&  \left.   + (1 -\tau)(\sqrt{(1 + \tau)/2}- 1) \,  K(\sqrt{(1 + \tau)/2})^3   \, \right]  , \nonumber   \end{align}
which diverges as $64/ [\pi^3 (1-\tau)]$ as $\tau \to 1$. 
\begin{figure}[h!]
\centering
\includegraphics[height=6cm]{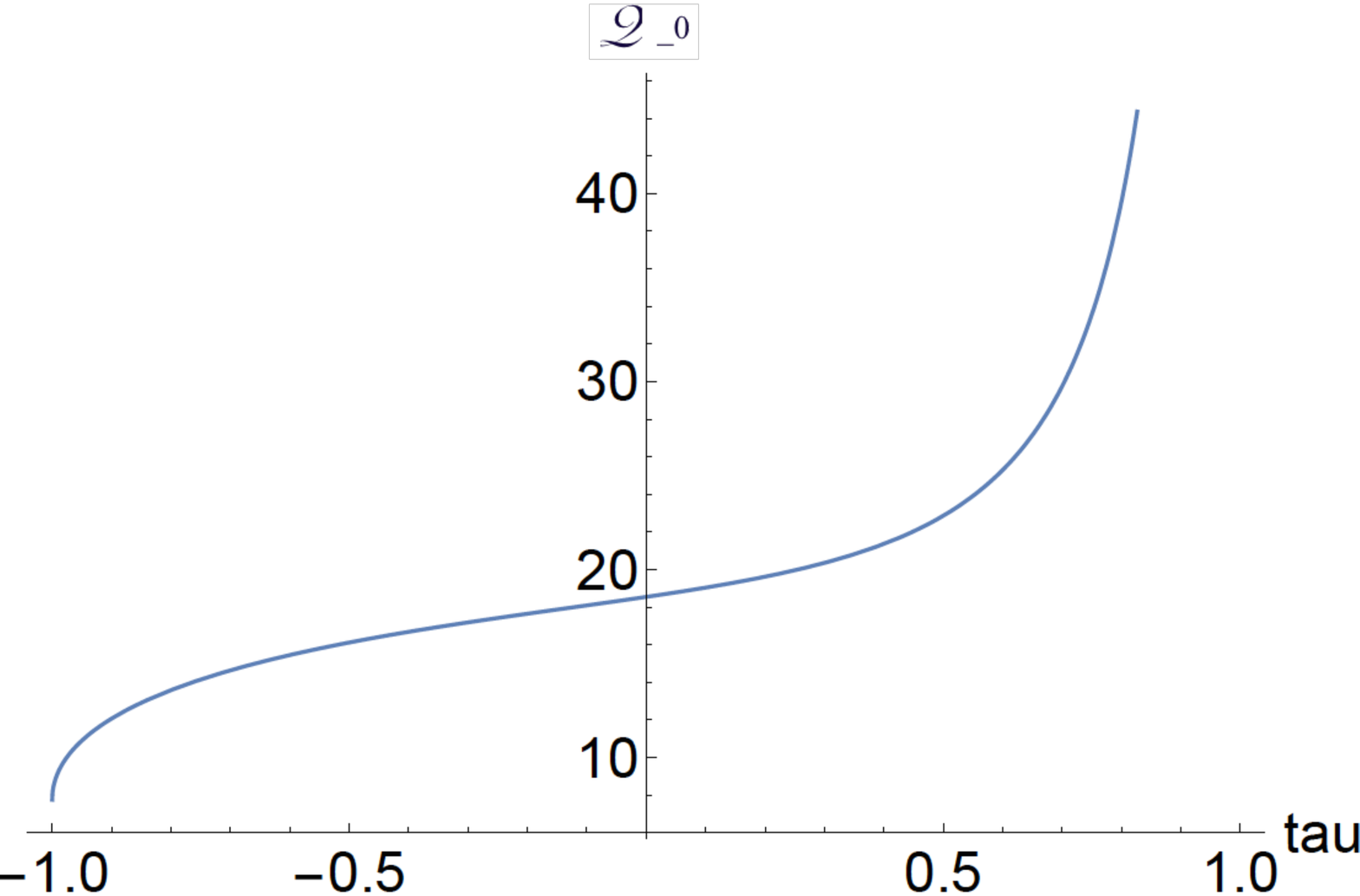}
\caption{Plot of $\pi^3 {{\cal  Q}_0} (\tau)/8$  as given by Eq.(\ref {explic}). 
}
\label{plotS}
\end{figure}
Similarly to Eqs.(\ref{8=}),  the differential operators in   Eqs.(\ref{pert2}) are singular at $\tau = \pm 1$. 
{ The  homogeneous solutions of Eq.(\ref{pert2}) for the electric field can be expressed as a linear  combination of $h_1(\tau)$ and $h_2(\tau)$ with 
\begin{align}\label{8*}
& h_1(\tau)= \frac{-1}{1 - \tau^2}\,  \left[(1 -\tau )\, K\left(\frac{1+\tau}{2}\right)  -2 E\left(\frac{1+\tau}{2}\right)\right] ,\\ & 
h_2(\tau)=  
\frac{-1}{1 - \tau^2}\, \left[ (1 +\tau)  \, K\left(\frac{1-\tau}{2}\right)  -2 E\left(\frac{1-\tau}{2}\right)\right ] , \nonumber  \end{align}
which diverge at $\tau = \pm 1 $ as   $1/(1 \mp \tau)$, respectively. 
\\
Analogously,  the  homogeneous solutions of Eq.(\ref{pert2}) for the magnetic  field can be expressed as a linear  combination of $m_1(\tau)$ and $m_2(\tau)$
\begin{align} \label{8**}
&  m_1(\tau) = \frac{1}{1 -\tau^2}  
\left[(1 -\tau) K\left(\frac{1+\tau}{2}\right) +2 \tau E\left(\frac{1+\tau}{2}\right)\right], \\
& m_2(\tau)  =  \frac{1}{1 -\tau^2}   \left[ (1 + \tau) K\left(\frac{1-\tau}{2}\right)  -2 \tau E\left(\frac{1-\tau}{2}\right) \right]. \nonumber \end{align}

The  inhomogeneous term  ${\cal Q}_0$ is  singular at $\tau = 1$. \, At  $\tau \to 1 $ the first two terms in Eq.(\ref{pert1b}) combine to give a contribution  proportional to
$ d [ (e_s + b_s)(e_s^2- b_s^2)]/d\tau$  that, according to Eq.(\ref{KQn1}) and accounting for cancellations, diverges as $ d\ln{(1-\tau)} /d\tau = 1/(1 - \tau)$
The third term in Eq.(\ref{pert1b}) diverges as $\ln^2{(1-\tau)}$.
The solution of Eq.(\ref{pert2}) that is initialised at $\tau =-1$ can be obtained with the general method of the variation of  the constants, in the form 
\begin{align} \label{9}
&e_{s1}(\tau) =  -  h_1({\tau})  \int_{-1}^\tau  \frac {h_2 (\tau')}{W(h_1,h_2)} \, \frac{1 }{1-\tau^2} \,  \frac {d  {{\cal Q}_0}(\tau')}{d \tau'}{d \tau'}\\ 
& ~ \qquad \quad + h_2({\tau}) \int_{-1}^\tau  \frac {h_1 (\tau')}{W(h_1,h_2)} \,  \frac{1 }{1-\tau^2} \,  \frac {d  {{\cal Q}_0}(\tau')}{d \tau'}{d \tau'} \nonumber \\
&~ \quad \quad  =  -  \frac{ 2 h_1({\tau})}{\pi}   \int_{-1}^\tau   {h_2 (\tau')} \, (1-\tau^2) \,  \frac {d  {{\cal Q}_0}(\tau')}{d \tau'}{d \tau'}  \nonumber \\ 
& ~ \qquad \quad + \frac{ 2 h_2({\tau})}{\pi}   \int_{-1}^\tau  {h_1 (\tau')} (1-\tau^2) \,  \frac {d  {{\cal Q}_0}(\tau')}{d \tau'}{d \tau'} ,  \nonumber 
\end{align}
where 
\begin{align} 
& W(h_1,h_2)= \frac{d h_1}{d \tau} h_2 - \frac{d h_2}{d \tau} h_1
= \frac{1} {(1-\tau^2)^2}  \left[   K\left(\frac{1-\tau}{2}\right) 
K\left(\frac{1+\tau}{2}\right)  \right. \nonumber  \\
&  \left. 
- K\left(\frac{1-\tau}{2}\right)  E\left(\frac{1+\tau}{2}\right) -K\left(\frac{1+\tau}{2}\right) E\left(\frac{1-\tau}{2}\right) 
 \right] \quad 
= \frac{\pi/2} {(1-\tau^2)^2}
\end{align}
is the Wronskian of the two independent homogeneous solutions
\begin{figure}[h!] \label{kernel}
\centering
\includegraphics[height=6cm]{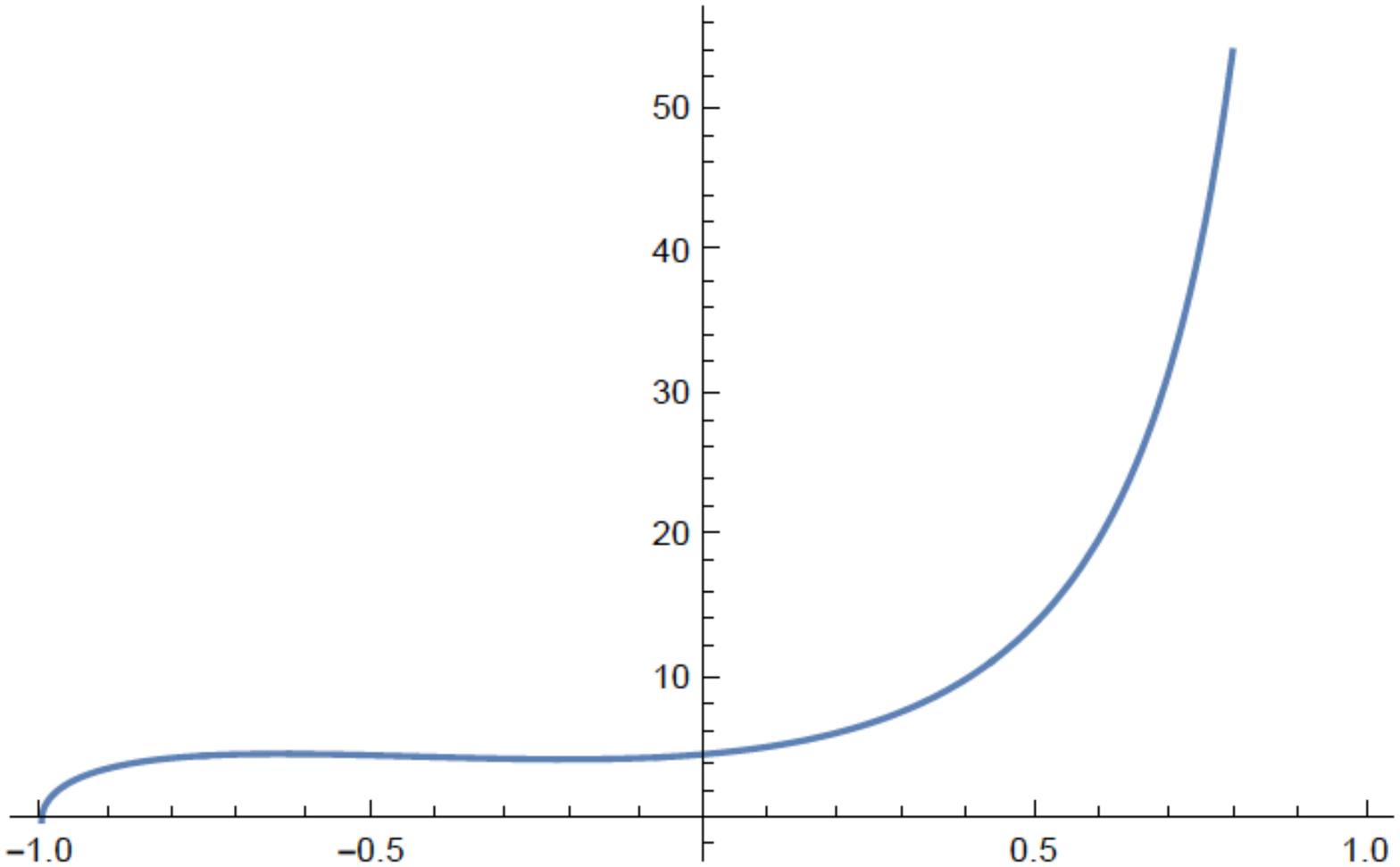}
\caption{Plot of $(\pi^3 /8) \, (1 - \tau^2)  \, d {{\cal Q}_0} (\tau) / d \tau$  as given by Eq.(\ref {explic}). }
\label{kern}
\end{figure}

The result of the numerical integration of Eq.(\ref{9}) is presented in Fig.(\ref{final})   where $\pi^3 e_{s1}(\tau)/8$ is plotted.
\begin{figure}[h!] \label{final}
\centering
\includegraphics[height=6cm]{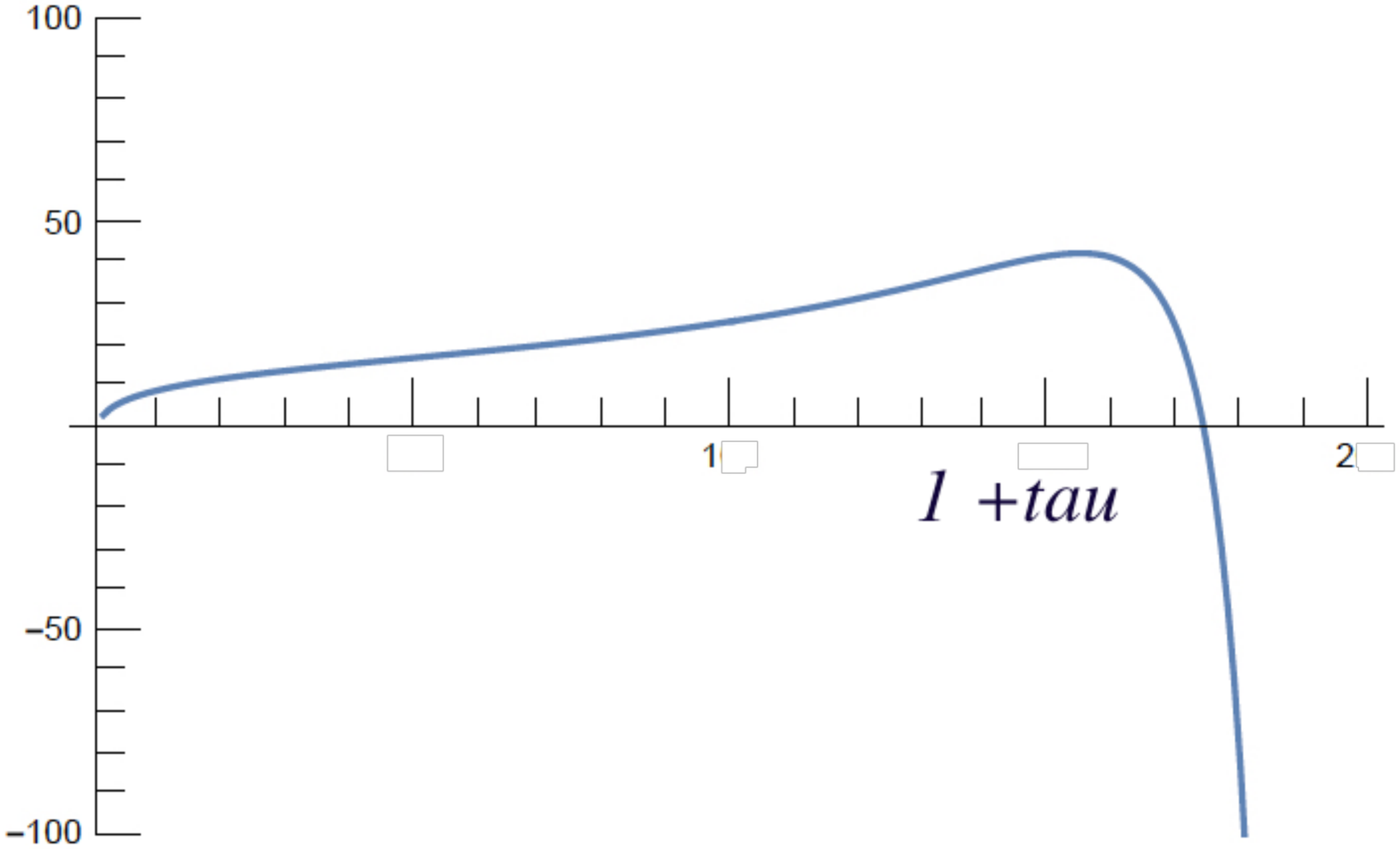}
\caption{Plot of $(\pi^3 /8) \, e_{s1}(\tau)$  showing a $1/(1-\tau)$ behaviour for $\tau \to 1$. }
\label{finalke}
\end{figure}}\\
Consistently with a local analysis based on Eqs.(\ref{pert2},\ref{explic}), Fig.(\ref{final})  shows   that $e_{s1}(\tau) $ diverges proportionally to $-1/(1 - \tau)$  as $\tau \to 1$.  \, Although formally the perturbation procedure breaks down at the singularity, the above result can be easily reinterpreted by observing that  the nonlinear interaction between the  converging and the diverging  portions of the electromagnetic  radiation  makes the radiation propagate at a speed smaller than the speed of light (see e.g. Ref.\cite{Gies}). This velocity reduction depends on the  field intensity. It is  thus proportional to the expansion parameter $\epsilon/r$ used  in Eq.(\ref{pert00}) and  leads to an $r$-dependent shift, for given $t$,  of the  position of the cumulation front. A similar  shift was noted  in Ref \cite{hod}  in the case of self-similar, counter-propagating waves in a Cartesian geometry. The expansion of the logarithmic singularity around the unperturbed 
  cumulation front   leads to the $- 1/(1 - \tau)$  divergent correction found above.  \,
  We note that, following the approach of Ref.\cite{hod}  this result  could have been derived more consistently by  using a renormalised expansion procedure where   a counter term is added to the leading order terms so as to make the first order correction finite, i.e.  by setting 
  \begin{equation} \label{renorm} 
  e_s (r,t)  = e_{s0}\left(\tau \left(1 - \frac{\epsilon }{r}v(\tau)\right)\right) + \frac{\epsilon}{r}  \, e_{s1}(\tau),\,  b_s = b_{s0}\left(\tau \left(1 - \frac{\epsilon }{r}v(\tau)\right)\right) + \frac{\epsilon}{r} \, b_{s1}(\tau), 
  \end{equation}
  and by solving for $v(\tau)$ so as to cancel the term in ${{\cal Q}_0}(\tau)$  that leads to  the $-1/(1 - \tau)$ singularity.

\section{Generation of electron positron pairs} \label{pair}

Two features  of the self-similar   converging and diverging electromagnetic fields in a cylindrical configuration can be exploited  for enhancing the production of  electron positron pairs.  \, First, as is the case for converging and focussed  beams  \cite{bulanino},    the Lorentz invariant ${\bf E}^2 - {\bf B}^2$  does not vanish and,  as shown by Eq.(\ref{Inv1}), it is positive for the $s$-polarisation i.e. for the case where the electric field is parallel to the cylinder axis while the magnetic field is azimuthal.  In addition,  at the cumulation front  $E^2 - B^2$ diverges logarithmically, thus formally exceeding the Schwinger field $E_S$.   On the other hand the width of enhanced field region is relatively narrow so that the pair production region will appear, at fixed $z$,   as a thin expanding ring. 
\\
For purely $s$-polarised fields the Lorentz invariant $ {\mathfrak G} ={\bf E}\cdot {\bf B}$ vanishes identically and  the pair production rate (see Ref. \cite{BLP}) is given
 { in dimensional units by
\begin{equation}\label{bulaninoform}
\frac{dN}{dt} = \frac{c}{4\pi^3} 
\left(\frac{m_ec}{\hbar}\right)^4
\int d V\, \left( \frac{E}{E_S}\right)^2  \exp{\left(-\frac{\pi E_S}{E}\right)}, 
\end{equation}
where $E$ is the invariant electric field  $E = (2  {\mathfrak F} )^{1/2} =  [2  ({\bf E}^2 - {\bf B}^2) ]^{1/2} $.  For the  $s$-polarisation 
the invariant electric field  $E$ in the interval $1 < \tau < 1$  normalised on the Schwinger field  $E_s$  is given by 
\begin{equation}\label{spolar}  \frac{E }{E_S}= \left(\frac{2}{r}\right)^{1/2} \left[\frac{4A_s}{\pi}\right] {\cal E}^{1/2}, \end{equation} 
where the factor $1/r$ arises from  the field representation in Eq.(\ref{3=}). A corresponding expression applies to the interval $ 1<\tau < \infty$.
In  Eq.(\ref{spolar})  $r$  is dimensional  while  the amplitude $A_s$ has the dimension of the square root of a length,  unlike from Eq.(\ref{Inv1}) where it is dimensionless. 
\,
Inserting Eq.(\ref{spolar}) into  the volume integral in Eq.(\ref{bulaninoform}) in cylindrical coordinates  we find 
\begin{equation}\label{EP2} 
\int d V  \left( \frac{E}{E_S}\right)^2  \exp{\left(-\frac{\pi E_S}{E}\right)} =
   4\pi \int  dz \int  dr \left[\frac{4A_s}{\pi}\right]^2 {\cal E}
\exp{\left( -\frac{\pi^2 (r/2)^{1/2}}{ 4  A_s {\cal E}^{1/2}} \right)}  .
\end{equation}
 At  fixed time $t$, the dominant contribution to the radial integral  arises from the neighbourhood of $ r = ct$  where 
${\cal E} \sim -(1/2) \ln{|1 -\tau|} \sim (1/2)|\ln{\xi}|$ with $\xi = | 1 - \tau |$.
Thus, taking into account the two sides of the logarithmic singularity (and using the continuity of the electromagnetic fields at $\tau = 1$)
we can rewrite the r.h.s. of  {Eq.(\ref{EP2})} as 
\begin{equation}\label{EP3} 
   4\pi  r   \left[\frac{4A_s}{\pi}\right]^2   \int  dz \int_0^{\bar \xi}  d\xi|\ln{\xi}| \,
\exp{\left( -\frac{\pi^2 \, r^{1/2}}{  4  \, A_s \, {|\ln{\xi}|}^{1/2}} \right)} ,
\end{equation}
where $r  = ct$ and the precise determination of the upper  limit  of integration ${\bar \xi}$ is not  needed since the integrand 
is strongly localised around $\xi = 0$ for most cases of interest.
 { Thus  the pair production rate  in Eq.(\ref{bulaninoform})   for the self-similar cylindrical $s$-polarised configuration  can be written per  unit length along   $z$} in the form   
\begin{align} \label{Ndot}  &
~ \quad \frac{1}{t}\frac{dN(t)}{dt} \sim 
\left(\frac{c}{\pi}\right)^2 \left(\frac{m_e c}{\hbar}\right)^4  {\cal A}^2 \,  {\cal R}(A)
\\ {\rm with} \qquad & {\cal R}({\cal A})= 
\int_0^{\bar \xi}  d\xi|\ln{\xi}| \,
\exp{\left( -\frac{\pi\, (ct)^{1/2}}{ {\cal A}\, {|\ln{\xi}|}^{1/2}} \right)}, \nonumber
\end{align}
where ${\cal A}= 4 A_s /\pi$.
\, The integral {${\cal R}({\cal A}) $ on the r.h.s. of Eq.(\ref{Ndot}) decreases extremely rapidly as ${\cal A}$ decreases. The plot of minus   the logarithm of ${\cal R}({\cal A})$ } as a function  of $ \pi (ct)^{1/2} / {\cal A}$  in Fig. \ref{positr}   shows   the rapid decrease of  the pair production rate with decreasing field amplitudes.
 \begin{figure}[h!] 
\centering
\includegraphics[height=6cm]{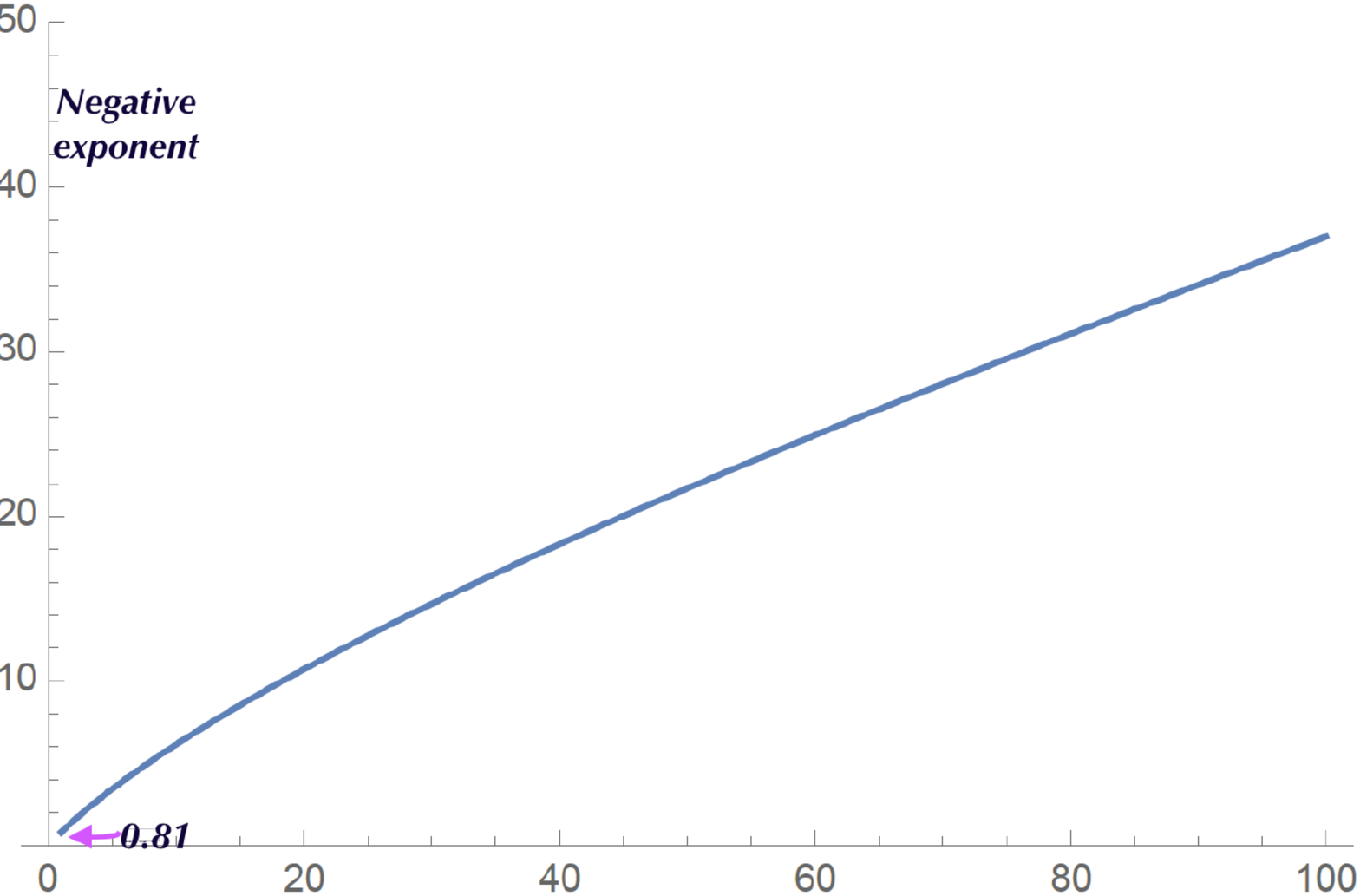}
\caption{Plot of $- \ln{{\cal R}({\cal A})}$ as a function of $ \pi  (ct)^{1/2}/ {\cal A})$. }
\label{positr}
\end{figure}
The total number $N_{tot}$ of electron-positron pairs per unit length along $z$ generated at the cumulation front by a self-similar  pulse initialised at $\tau= -1$ can be estimated by taking the integral over time in Eq.(\ref{Ndot}) first and  then by performing the integral over $\xi$.
\,
Using  the relationship 
$ \int_0^{+\infty}  t \exp{(- a t^{1/2})} \, dt =  12/a^4 $
we obtain  
\begin{equation} \label{Ntotapp} 
N_{tot} \sim 
\left(\frac{c}{\pi}\right)^2 \left(\frac{m_e c}{\hbar}\right)^4  \frac{12}{ c^2 \pi^4} {\cal A}^6 \,  
\int_0^{\bar \xi}  d\xi |\ln{\xi}|^3 .
\end{equation}
Then, taking ${\bar \xi} = 1$ and using
$ \int_0^{1} d\xi |\ln{\xi}|^3 = 6, $ we find
\begin{equation} \label{Ntot} 
N_{tot} \sim 
\left(\frac{m_e c}{\hbar}\right)^4  \frac{72}{\pi^6} {\cal A}^6 \,  =N_0
 \left(\frac{m_e c}{\hbar}\right)^4  A_s^6, \quad  {\rm with} \,N_0 = 9 \left(
\frac {2^{1/2}\, 4}{\pi^2}\right)^6 \approx 0.32 . \end{equation}
While the exact value of the numerical coefficient $N_0 $ can be affected by the approximations made in its calculation,  the dependence on  $A_s^6$  
can be understood  simply from  first principles.   First we observe that the number $N_{tot}$ of electron-positron pairs per unit length has the dimension of an inverse length while ${m_e c}/{\hbar}$ is the inverse of the reduced Compton length ${\lambdabar} $.  By definition a self-similar configuration cannot provide an intrinsic scale-length  to be used in the definition of the volume that is needed to balance Eq.(\ref{Ntot}) dimensionally.  However a spatial scale can be derived from the behaviour of the  electromagnetic fields in a cylindrical configuration as given by Eq.(\ref{3=}). In fact the amplitude $A_s$  has the dimension  of the square root of a length and $A_s^2$ can be read  as the radial distance $r_S$  from the axis where the electric field amplitude is equal to the Schwinger field.  Thus Eq.(\ref{Ntot})  can be  interpreted  by saying that the number of pairs produced in a disc of height  along $z$ equal to $\lambdabar$  is given by a numerical coefficient  of order unity times the  cube of the ratio  $r_S/\lambdabar$.

  \section{Conclusions} \label{concl}
  
In this article we have analysed  two QED  effects  on a properly arranged, cylindrical,  electromagnetic configuration, with a high degree of spectral coherence, that develops an expanding cumulation front where the electromagnetic  fields diverge logarithmically. 
  We have shown that QED effects make the reflected cumulation front  {expand} with a velocity smaller  than the speed of light, similarly to what is known to occur  when two  counter-propagating  electromagnetic pulses interact nonlinearly. We have computed the effect of the enhancement of the electromagnetic fields at the expanding front on the production rate of electron-positron pairs.  {We have shown that the total number of pairs produced scales as the sixth power of the electromagnetic field amplitude.}

 The analysis presented above is not exhaustive in several aspects. One concerns the realisability in the laboratory of the highly coherent fields that lead to the cumulation process and  the assessment of  the resilience of the cumulation mechanism to errors in the field generation.  Moreover  it may not be fully consistent  to  use the  Euler Heisenberg Lagrangian,  which is derived in the long wavelength limit, to account for the QED effect too close to the cumulation front {where the local gradient of the field amplitude diverges as $1/|1-\tau | $}.
 \\ {Two obvious problems arise from the slow decay  of the frequency spectrum as $\omega \to \infty$ and from the proper boundary conditions to be  imposed to a ``local''  realisation of  the self-similar solution in order to obtain  a finite duration pulse.We have  shown that, if we impose e.g. a frequency  Gaussian cut-off  with  width $\omega_{max}$  in the frequency spectrum while preserving the spectrum coherence,  the  amplitude cumulation turns out to be   bounded and that  the logarithmic singularity at $\tau =1$  is changed into  $e_s(\tau = 1) \propto \ln{(\omega_{max} r/c)^2}$. \\ Regarding the pulse duration, a finite  pulse that exhibits amplitude  cumulation,  although transiently in time, can be obtained by considering a ``truncated'' self-similar solution. In the derivation in Sec.\ref{s-sim} the self-similar  fields  are initiated at $\tau = - 1$  that is for all  values of $r $ and, for each $r$,  at the  corresponding  past  time $ t = - r/c$.  We may consider instead a truncated  self-similar solution,  that is  a solution that is initiated at $\tau = -1$ but extends only on a finite $r$ interval, and thus on a finite $t$ interval.
\\
Since all parts of the self-similar  solutions  propagate at the same speed of light $c$  causality requires  that  this ``truncated ''  part of the self-similar  solution propagates at $c$  and if it is sufficiently long, i.e. more than two times the distance from the cylinder axis,  for a  finite time interval  it will not suffer from the fact that portion of the self-similar solutions are missing leading to a transient cumulation front.  Clearly  this condition must be satisfied in the interval where  the constructive interference with the reflected pulse can occur which can substantially reduce the portion of the truncated pulse that can be used for building the transient cumulation front, thus reducing the efficiency of the process.

The  formation of the  singularity  during the  cumulation of  a strong electromagnetic wave  formally  assumes that the frequency spectrum of the wave does not decay exponentially when  the frequency tends to infinity,  as discussed above. The counter-play between the electromagnetic field intensification and the wave steepening at the singularity may result either in stronger electron-positron pair generation by the Schwinger effect, as considered above, or in the  creation  of a pair plasma because  the Breit-Wheeler mechanism becomes dominant (see the review article \cite{piazza} and references therein) and/or in the modification of  the singularity   due to dispersion effects.

We note that the frequency spectrum of the electromagnetic wave can be determined not only by the boundary conditions but also 
by the high order harmonics generated in the nonlinear vacuum. These  high order harmonics  are  a manifestation of nonlinear processes and  have attracted a lot of attention in theoretical articles devoted to the study of the QED vacuum \cite{DHK-2005,BKR-2015,FN-2007,KKB-2019,SEPB-2020}. 
The harmonic development can modify the singularity  formed in the QED vacuum. In combination with the dispersion 
effects, the counter-play between nonlinearity and dispersion can result in the  formation of solitons \cite{BSPKBER-2020}. 
This regime is beyond  the scope of the present work and will be addressed in a future paper.

We conclude by stating that, notwithstanding these limitations,  our analysis indicates that it may be worthwhile to reconsider, e.g. in a three-dimensional configuration,  the process of amplitude cumulatiosn described in a cylindrical geometry  in Ref.\cite{cumul}   as it may represent a promising  new  approach to the study of QED effects in the laboratory. }

\section*{Acknowledgements}

S.V.B. acknowledges the support by the project High Field Initiative\\
 (CZ.02.1.01/0.0/0.0/15\_003/0000449) from the European Regional Development Fund.   
 \\
~
\\
F.P. would like to acknowledge  the hospitality of the ELI-Beamlines Project, Na Slovance 2, 182 21 Prague, Czech Republic.

\end{document}